\documentclass[twocolumn,apjl]{emulateapj}
\usepackage{apjfonts}
\epsscale{1.15}

\shorttitle{A Kpc-Scale Binary AGN} 
\shortauthors{Fu et al.}
\journalinfo{accepted by ApJL}
\submitted{accepted by ApJL}

\newcommand{\chandra}{{\it Chandra}}

\newcommand{\iras}{{\it IRAS}}
\newcommand{\rosat}{{\it ROSAT}}

\newcommand{\fir}{{FIRST}}
\newcommand{\kms}{{km s$^{-1}$}}
\newcommand{\msun}{$M_{\odot}$}

\newcommand{\um}{$\mu$m}
\newcommand{\uJy}{$\mu$Jy}

\newcommand{\Ha}{H$\alpha$}
\newcommand{\Hb}{H$\beta$}
\newcommand{\OII}{[O\,{\sc ii}]}
\newcommand{\OIIexpanded}{[O\,{\sc ii}]\,$\lambda$3727}

\newcommand{\OIII}{[O\,{\sc iii}]}
\newcommand{\OIIIexpanded}{[O\,{\sc iii}]\,$\lambda$5007}

\newcommand{\MgII}{Mg\,{\sc ii}}
\newcommand{\objfull}{SDSS\,J150243.1$+$111557}
\newcommand{\obj}{SDSS\,J1502+1115}
\newcommand{\sersic}{S\'{e}rsic}

\begin{document}

\title{A Kiloparsec-Scale Binary Active Galactic Nucleus Confirmed by the Expanded Very Large Array}

\author{Hai Fu\altaffilmark{1}, Zhi-Yu Zhang\altaffilmark{2,3,4}, Roberto J. Assef\altaffilmark{5}, Alan Stockton\altaffilmark{6}, Adam D. Myers\altaffilmark{7,8}, Lin Yan\altaffilmark{9}, S. G. Djorgovski\altaffilmark{1,10}, J. M. Wrobel\altaffilmark{11}, and Dominik A. Riechers\altaffilmark{1}
}
\altaffiltext{1}{Astronomy Department, California Institute of Technology, MS 249$-$17, Pasadena, CA 91125, USA} 
\altaffiltext{2}{Purple Mountain Observatory, CAS, 2 West Beijing Road, Nanjing, 210008, China}
\altaffiltext{3}{Max-Planck-Institut f\"ur Radioastronomie, Auf dem H\"ugel 69, D-53121 Bonn, Germany}
\altaffiltext{4}{Graduate School of the Chinese Academy of Sciences, 19A Yuquan Road, P.O. Box 3908, Beijing 100039, China}
\altaffiltext{5}{NASA Postdoctoral Program Fellow at the Jet Propulsion Laboratory, California Institute of Technology, MS 169-530, 4800 Oak Grove Dr., Pasadena, CA 91109, USA}
\altaffiltext{6}{Institute for Astronomy, University of Hawaii, 2680 Woodlawn Drive, Honolulu, HI 96822, USA}
\altaffiltext{7}{Department of Physics and Astronomy, University of Wyoming, Laramie, WY 82071, USA}
\altaffiltext{8}{Max-Planck-Institut f\"ur Astronomie, K\"onigstuhl 17, D-69117 Heidelberg, Germany}
\altaffiltext{9}{Spitzer Science Center, California Institute of Technology, MS 220$-$06, Pasadena, CA 91125, USA} 
\altaffiltext{10}{Distinguished Visiting Professor, King Abdulaziz University, Jeddah, Saudi Arabia}
\altaffiltext{11}{National Radio Astronomy Observatory, P.O. Box O, Socorro, NM 87801, USA}

\begin{abstract} 
We report the confirmation of a kpc-scale binary active galactic nucleus (AGN) with high-resolution radio images from the Expanded Very Large Array (EVLA). \objfull\ is a double-peaked \OIII\ AGN at $z = 0.39$ selected from the Sloan Digital Sky Survey. Our previous near-infrared adaptive optics imaging reveals two nuclei separated by 1.4\arcsec\ (7.4~kpc), and our optical integral-field spectroscopy suggests that they are a type-1$-$type-2 AGN pair. However, these data alone cannot rule out the single AGN scenario where the narrow emission-line region associated with the secondary is photoionized by the broad-line AGN in the primary. Our new EVLA images at 1.4, 5.0, and 8.5~GHz show two steep-spectrum compact radio sources spatially coincident with the optical nuclei. The radio power of the type-2 AGN is an order-of-magnitude in excess of star-forming galaxies with similar extinction-corrected \OIIexpanded\ luminosities, indicating that the radio emission is powered by accretion. Therefore, \objfull\ is one of the few confirmed kpc-scale binary AGN systems. Spectral-energy-distribution modeling shows that \objfull\ is a merger of two $\sim10^{11}$ \msun\ galaxies. With both black hole masses around $10^8$ \msun, the AGNs are accreting at $\sim10$ times below the Eddington limit.
\end{abstract}


\section{Introduction} \label{sec:introduction}

Binary supermassive black holes (SMBHs) resulting from galaxy mergers should be commonplace, given that virtually every massive galaxy harbors a SMBH \citep{Richstone98}. However, cases are rare where both SMBHs in a merging system are accreting as active galactic nuclei (AGNs). Hundreds of binary AGNs with $>$10~kpc separations have been discovered both at high redshifts \citep{Myers08,Hennawi10} and locally \citep{Liu11a}. But, there are only a few \emph{confirmed} kpc-scale binary AGNs \citep{Junkkarinen01,Komossa03,Koss11}. Such close binaries are important, because strong tidal interactions are more likely to influence the nuclear accretion and star formation in both galaxies at kpc-scale separations \citep{Barnes96}.

Systematic searches for kpc-scale binary AGNs have focused on AGNs with pairs of narrow \OIIIexpanded\ emission lines in spatially integrated spectra \citep{Gerke07,Comerford09a,Wang09,Liu10a,Smith10}, which account for $\sim$1\% of the AGNs in the Sloan Digital Sky Survey (SDSS). SDSS data are inadequate to confirm such binaries because they are typically spatially unresolved in seeing-limited optical data. However, subsequent high-resolution imaging and spatially resolved spectroscopy have found that most of these AGNs show double-peaked line profiles because of gas kinematics driven by outflows, inflows, or disk rotation \citep{Shen11,Fu11b}. Nonetheless, these observations have also found a few binary AGN candidates that show Seyfert-like emission-line regions that spatially coincide with the stellar nuclei of the merging galaxies \citep{Liu10b,McGurk11,Fu11b}. Yet, none of these candidates has been confirmed to host two actively accreting SMBHs, because with the optical data alone, it is difficult to rule out that a Seyfert-2-type emission-line region is photoionized by an AGN \emph{outside} of that region. High-resolution radio or X-ray detections of both nuclei can provide unambiguous evidence for active accretion (see the VLBA searches of \citealt{Tingay11} and \citealt{Burke11}). 

\objfull\ ($z = 0.39$, $r = 18.0$; hereafter \obj) is a double-peaked type-1 AGN identified by \citet{Smith10}. Our previous observations suggest that it is a seven-kpc-separation type-1$-$type-2 AGN pair \citep{Fu11b}. Here we report radio continuum imaging from the Expanded Very Large Array \citep[EVLA;][]{Perley11}, which confirms \obj\ as a binary AGN. Throughout we adopt a $\Lambda$CDM cosmology with $\Omega_{\rm m}=0.3$, $\Omega_\Lambda=0.7$ and H$_0$ = 70 km~s$^{-1}$~Mpc$^{-1}$. 

\section{Observations} \label{sec:obs}

\begin{figure*}[!t]
\plotone{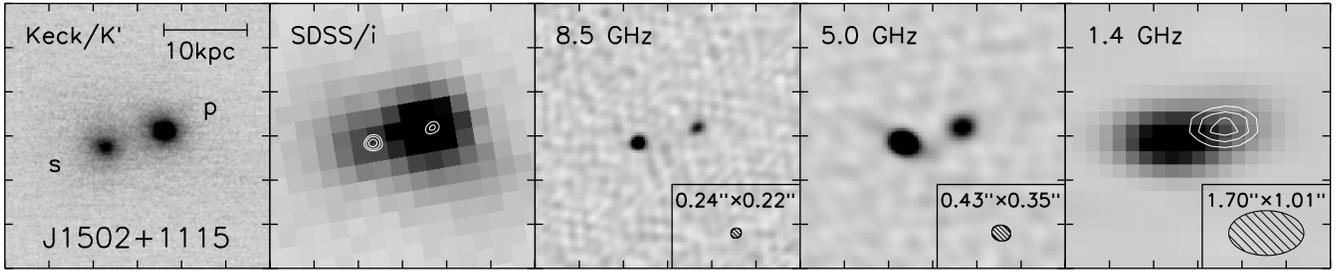}
\caption{Keck/$K'$-band, SDSS/$i$-band, and EVLA/$X-C-L$-bands 6\arcsec\ image stamps of \obj. The scale bar in the $K'$ image indicates a transverse separation of 10~kpc. The primary (``p'') and secondary (``s'') AGNs are labeled. The 8.5~GHz contours in the $i$-band image are at 8, 16, 32-$\sigma$, where 1-$\sigma$ = 23 \uJy~beam$^{-1}$. To better illustrate the primary in the 1.4~GHz panel, we show contours from the residual image after subtracting the secondary. The contour levels are 32, 40, and 48-$\sigma$, where 1-$\sigma =$ 44 \uJy~beam$^{-1}$. Hatched ellipses at the lower right corners of the radio maps show the EVLA beams. The images are aligned based on their own astrometry and are displayed in linear scale. N is up and E is to the left. Tickmarks are spaced in 1\arcsec. 
\label{fig:allimage}}
\end{figure*}

We observed \obj\ with the EVLA in the A configuration under project code 11A-175. The EVLA was set up with dual circular polarizations and a total bandwidth of 256~MHz per polarization at central frequencies of 1.391, 4.959, and 8.459~GHz (hereafter 1.4, 5.0, 8.5~GHz). The 1.4 and 5.0~GHz observations were carried out consecutively on 2011 June 29 (UT), and the 8.5~GHz observations were carried out on 2011 June 14 (UT). The total on-source exposure times were 25, 35, and 20~minutes at 1.4, 5.0, and 8.5~GHz, respectively. The nearby point-like quasar J1504+1029 was used as the gain calibrator. The switching angle was 53\arcmin, and the switching times were 7, 10, and 10 minutes at 1.4, 5.0, and 8.5~GHz, respectively. 3C~286 was observed for bandpass and flux-density calibration.

Data calibration was carried out with the Common Astronomical Software Applications (CASA) package, while imaging employed both CASA and AIPS (Astronomical Image Processing System). For the 5.0 and 8.5~GHz data, we used a Briggs robust parameter of $0$ to achieve good image fidelity with only a modest loss of sensitivity compared to ``natural'' weighting. For the 1.4~GHz data, we used ``uniform'' weighting to achieve the best resolution. The final images, shown in Fig.~\ref{fig:allimage}, have rms noise levels of 44, 17, and 23 $\mu$Jy beam$^{-1}$ for 1.4, 5.0, and 8.5~GHz, respectively.

\obj\ is resolved into two components, a primary (``p'') and a secondary (``s''), separated by 1.4\arcsec\ at all frequencies (Fig.~\ref{fig:allimage}). Their 8.5~GHz centroids are at RA = 15:02:43.088, Dec = $+$11:15:57.42 (p), and RA = 15:02:43.179, Dec = $+$11:15:57.07 (s; J2000). The secondary shows hints of adjacent structure at 5.0~GHz; to quantify this structure at 8.5~GHz, a matched-resolution image was made by excluding visibility data with baselines longer than those available at 5.0~GHz. This matched-resolution image was used for component photometry at 8.5~GHz.

The integrated flux densities of the primary and secondary at each frequency appear in Table 1, where the errors are the quadratic sum of the error in the fit and a 3\% error in the flux density scale. The combined 1.4 GHz flux density is $9.62\pm0.42$ mJy, consistent with the integrated flux from the \fir\ catalog \citep[$9.30\pm0.49$ mJy;][]{White97}. Both sources are compact. The beam-deconvolved sizes of both components are less than 0.3\arcsec\ (1.6~kpc) at both 5.0 and 8.5~GHz; the size measurements from the 1.4~GHz image are less accurate.

\section{Analysis and Results} \label{sec:results}

\begin{figure*}[!t]
\plotone{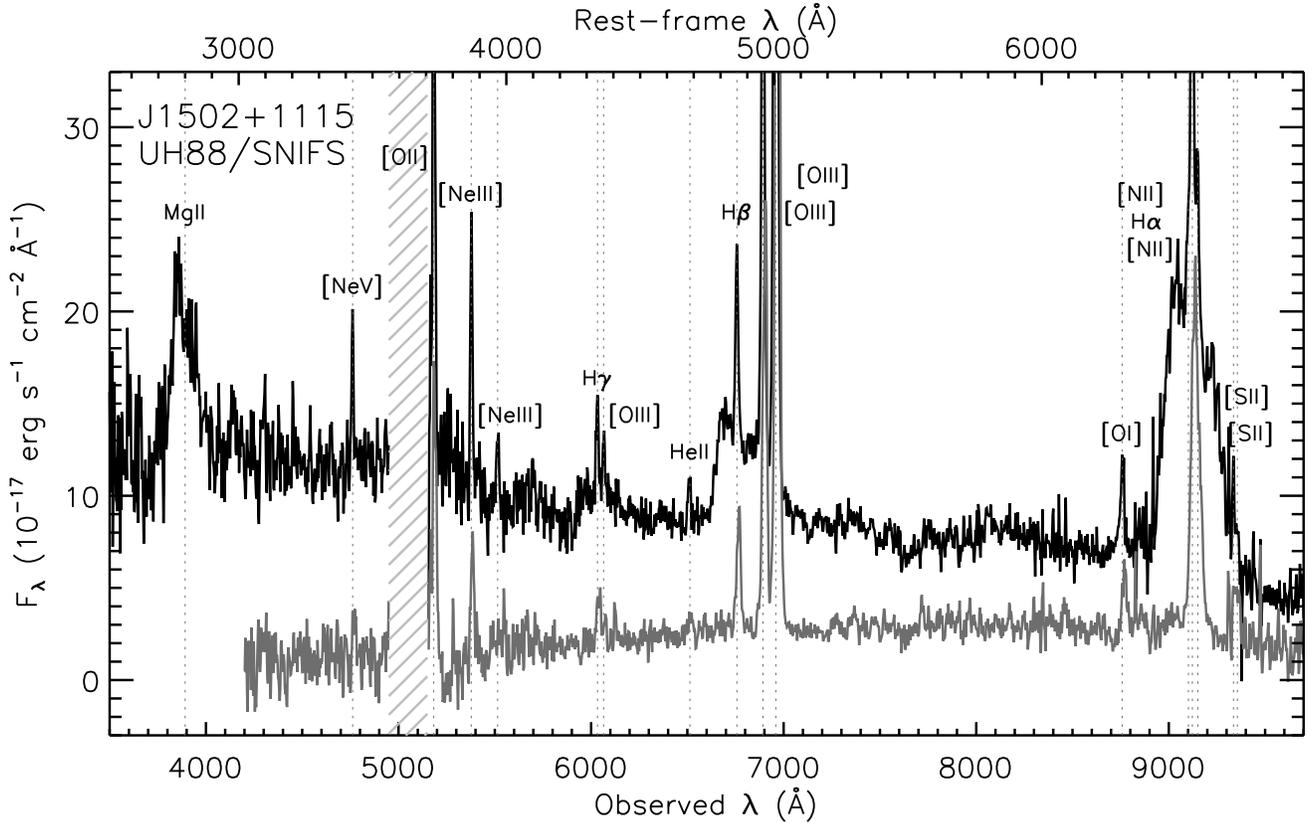}
\caption{Optical spectra of the two active nuclei in \obj. The spectra were extracted from the SNIFS datacube using 0.8\arcsec\ diameter apertures. The primary nucleus is shown in black, and the secondary is in grey. We have labeled strong emission lines and hatched the gap between the blue and red channels. Because of atmospheric refraction, the secondary shifts out of the datacube below $\sim$4200 \AA.
\label{fig:snifs}}
\end{figure*}

The EVLA images are analyzed in combination with a $K'$-band image from the Keck\,II laser-guide-star adaptive optics system \citep{Wizinowich06} and seeing-limited optical integral-field spectra from the Supernova Integral-Field Spectrograph \citep[SNIFS;][]{Aldering02} on the UH 2.2-meter telescope. The $K'$-band image resolved \obj\ into two components separated by 1.39\arcsec\ (7.4~kpc), and the integral-field spectroscopy spatially resolved the kinematic components and associated them with the two stellar nuclei \citep[][]{Fu11b}. The \OIII\ line of the primary is blueshifted relative to that of the secondary by 657 \kms. In Figure~\ref{fig:snifs}, we show the optical spectra of the two components extracted from the SNIFS datacube with 0.8\arcsec\ apertures. The primary is a type-1 AGN with broad \MgII\ and Balmer emission lines along with high-ionization narrow lines, while the secondary is a type-2 AGN with narrow high-ionization lines and line ratios that are consistent with AGN photoionization according to the \citet{Baldwin81} diagrams \citep{Fu11b}. 

We compare the EVLA images with the SDSS $i$-band and the $K'$-band images in Figure~\ref{fig:allimage}. The offset between the two radio sources and the near-infrared (NIR) sources is only 0.1\arcsec, much smaller than the 1-$\sigma$ dispersion ($\sim$0.4\arcsec) between typical SDSS and \fir\ positions. 

Both sources show steep radio spectra, with $\alpha$ = $-0.80\pm0.06$ and $-0.92\pm0.04$ ($F_{\nu} \propto {\nu}^{\alpha}$) for the primary and secondary, respectively, indicating optically thin synchrotron emission. Compact steep-spectrum radio sources like \obj\ make up $\sim$30\% of bright radio sources and they are mostly powerful AGNs \citep{ODea98}. Since star formation and black hole accretion are the two major powering sources of extragalactic synchrotron radiation, we consider the two scenarios in the following.

\subsection{Nature of the Radio Emission in \obj} \label{sec:sf_vs_agn}

If the radio emission is entirely driven by star formation, the rest-frame 1.4~GHz power of $4.6\times10^{23}$ and $1.3\times10^{24}$ W~Hz$^{-1}$ imply star formation rates (SFRs) of 273 and 787 \msun~yr$^{-1}$ for the primary and secondary, respectively \citep{Yun01}. However, these tremendous amounts of star formation disagree with other well-established star formation indicators, such as the the far-infrared (FIR) luminosity and the \OIIexpanded\ luminosity, especially for the type-2 secondary, indicating that star formation is unlikely to dominate the radio emission.

First, \obj\ is undetected by \iras\ in all four bands (12, 25, 60, and 100 \um). The $scanpi$\footnote{http://scanpi.ipac.caltech.edu:9000/applications/Scanpi} 60\um\ 3-$\sigma$ upper limit of 0.15~Jy sets an upper limit of $2.2\times10^{23}$ W~Hz$^{-1}$ to the 1.4~GHz power from star formation, adopting the FIR-radio correlation of \citet{Yun01}. This total upper limit is 2.0 and 5.9 times \emph{less} than observed for the primary and secondary alone.

Secondly, the \OII\ luminosities give SFRs much less than those from the radio luminosity, even if we make the unrealistic assumption that all of the \OII\ emission comes from star formation. We use the narrow \Ha/\Hb\ ratio to estimate the intrinsic extinction, given an intrinsic ratio of 3.1 \citep{Osterbrock06}. We find extinctions of $A_V^p =$ 0.78 and $A_V^s = 0.28$. We then use \citet{Cardelli89} extinction curve to deredden the total \OII\ fluxes\footnote{These have been corrected for Galactic extinction and aperture loss.} ($f_{\rm [O~II]}^p = 1.0\times10^{-14}$ and $f_{\rm [O~II]}^s = 7.6\times10^{-15}$ erg~s$^{-1}$~cm$^{-2}$). The dereddened luminosities imply SFRs of 82 and 31 \msun~yr$^{-1}$ \citep{Kennicutt98}, which are 3.3 and 26 times less than the SFRs from the 1.4~GHz power for the primary and secondary, respectively.

To explore whether the radio emission can be powered by black hole accretion, we can consider whether the radio emission is in significant excess of the X-ray emission when compared with typical AGNs. \obj\ is detected by \rosat\ with an X-ray flux of $S_{\rm 0.1-2.4~keV} = (4.9\pm1.7)\times10^{-13}$ erg~cm$^{-2}$~s$^{-1}$ ($0.0429\pm0.0147$ count~s$^{-1}$). Accounting for the Galactic absorption ($N_{\rm H} = 2.37\times10^{20}$ cm$^{-2}$), we estimated\footnote{webPIMMS: http://heasarc.nasa.gov/Tools/w3pimms.html} rest-frame X-ray luminosities of $L_{\rm 0.5-2~keV} = (6.5\pm2.2)\times10^{43}$ erg~s$^{-1}$ and $L_{\rm 2-10~keV} = (1.0\pm0.6)\times10^{44}$ erg~s$^{-1}$ for a photon index $\Gamma = 1.8$. This X-ray power and the absolute rest-frame $B$-band magnitude $M_{\rm B} = -22.7$ place \obj\ on the borderline between Seyferts and quasars \citep{Brusa07}. The rest-frame 5~GHz luminosity of the entire system is $3.0\times10^{40}$ erg~s$^{-1}$. The 5~GHz to hard X-ray luminosity ratio, log$R_{\rm X}$ = log($\nu L_{\rm 5~GHz}/\nu L_{\rm 2-10~keV}$) = $-3.5$, lies between the median values of Seyfert-1s (log$R_{\rm X}$ = $-3.8$) and Seyfert-2s \citep[$-3.3$; see table 1 of][]{Ho08}. Therefore, most of the radio emission could be powered by accretion. The intermediate log$R_{\rm X}$ value also hints that \obj\ is a type-1$-$type-2 binary AGN.

\subsection{Spectral Energy Distribution Modeling} \label{sec:sed}

\begin{figure*}
\plotone{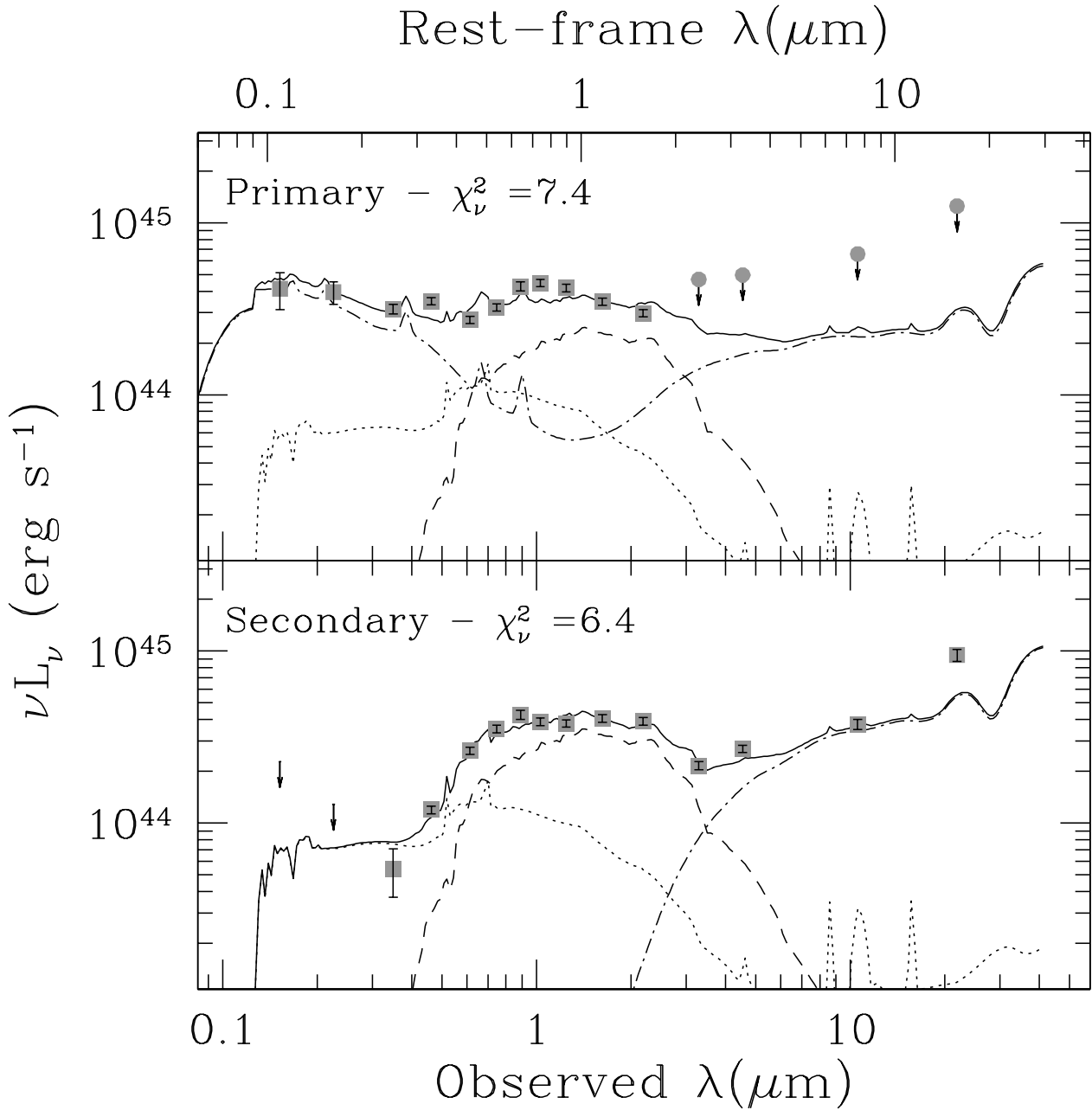}
\caption{Best-fit SED ({\it{solid line}}) to the primary ({\it{top}}) and secondary ({\it{bottom}}) components of \obj\ using the templates of \citet{Assef10}. The best-fit SEDs are a combination of a Type-1 AGN ({\it{dot-dashed line}}) with an old stellar population ({\it{dashed line}}) and a star-forming component ({\it{dotted line}}). The photometry of the primary and secondary components are shown by gray solid squares. WISE fluxes could not be deblended and are assumed as upper bounds for the primary component SED fit ({\it{gray solid circles}}). See \S~\ref{sec:sed} for details.}
\label{fg:sed_fit}
\end{figure*}

\begin{deluxetable}{cccc} 
\tablewidth{0pt}
\tablecaption{Spectral Energy Distribution
\label{tab:photo}}
\tablehead{ 
\colhead{Band} & \colhead{$\lambda$} & \colhead{$F_{\nu}$(Primary)} & \colhead{$F_{\nu}$(Secondary)} \\
\colhead{} & \colhead{(\um)} & \colhead{(mJy)} & \colhead{(mJy)} \\
\colhead{(1)} & \colhead{(2)} & \colhead{(3)} & \colhead{(4)}  
}
\startdata
    FUV &      0.16& \multicolumn{2}{c}{ 0.043$\pm$0.010} \nl
    NUV &      0.23& \multicolumn{2}{c}{ 0.061$\pm$0.009} \nl
    $u$ &      0.36& 0.075$\pm$0.005& 0.012$\pm$0.005 \\
    $g$ &      0.47& 0.110$\pm$0.003& 0.038$\pm$0.003 \\
    $r$ &      0.62& 0.114$\pm$0.005& 0.110$\pm$0.005 \\
    $i$ &      0.75& 0.164$\pm$0.026& 0.179$\pm$0.026 \\
    $z$ &      0.89& 0.259$\pm$0.016& 0.258$\pm$0.016 \\
    $Y$ &      1.03& 0.313$\pm$0.006& 0.272$\pm$0.006 \\
    $J$ &      1.25& 0.354$\pm$0.006& 0.321$\pm$0.006 \\
    $H$ &      1.63& 0.385$\pm$0.010& 0.448$\pm$0.010 \\
    $K$ &      2.20& 0.444$\pm$0.012& 0.582$\pm$0.012 \\
  WISE1 &      3.35& \multicolumn{2}{c}{ 1.049$\pm$0.026} \nl
  WISE2 &      4.60& \multicolumn{2}{c}{ 1.538$\pm$0.038} \nl
  WISE3 &     11.56& \multicolumn{2}{c}{ 4.733$\pm$0.179} \nl
  WISE4 &     22.09& \multicolumn{2}{c}{18.688$\pm$1.102} \nl
8.5 GHz &  35290.46&  0.61$\pm$0.04 &  1.28$\pm$0.06 \\
5.0 GHz &  60454.21&  0.93$\pm$0.05 &  2.33$\pm$0.08 \\
1.4 GHz & 215522.97&  2.58$\pm$0.25 &  7.04$\pm$0.34 \\
\enddata
\tablecomments{
Because we cannot separate the two components in GALEX and WISE images, we list their total fluxes.
}
\end{deluxetable}

We have compiled 15-band photometry from GALEX \citep[FUV, NUV;][]{Martin05}, SDSS DR8 \citep[$u$, $g$, $r$, $i$ and $z$;][]{Aihara11}, UKIRT Infrared Sky Survey (UKIDSS) DR6plus \citep[$Y$, $J$, $H$ and $K$;][]{Lawrence07}, and WISE \citep[3.4, 4.6, 12, and 22\um;][]{Wright10} surveys. The primary and secondary are blended in all of these catalogs but can be deblended in the SDSS and UKIDSS images. We use GALFIT \citep[][]{Peng10} to deblend the SDSS and UKIDSS images. We cannot separate the two components in the GALEX and WISE images given the small separation and their large point-spread-functions (PSFs). For the GALFIT input PSF, we use a stellar source 83\arcsec\ to the southwest. For the primary, either a direct PSF or a slightly Gaussian-convolved (FWHM $\sim$ 1 pixel = 0.4\arcsec) PSF provides a good fit. For the secondary, we used a \sersic\ profile. While modeling the $g, r$, and $i$-band images, we also masked out areas between the two galaxies and south of the primary that are contaminated by \OII\ and \OIII\ emission lines from the extended nebulae. For the $u, g, r$, and $z$-band images where the two components are not as well resolved as in the other bands, we fixed the offset between the two components to that determined from the Keck AO image. The deblended $K$-band fluxes from the UKIDSS image agree within 1-$\sigma$ with those measured from the AO image. We measured all fluxes with a 3\arcsec-radius aperture centered between the primary and the secondary. We chose this aperture because it includes most of the flux and it agrees with the SDSS $r$-band Petrosian radius. The deblended photometry is tabulated in Table~\ref{tab:photo}.

We fit the photometry using the empirical UV-to-mid-IR Spectral Energy Distribution (SED) templates of \citet{Assef10}. Every object is modeled by a non-negative linear combination of three galaxy templates, corresponding to an old stellar population, a moderately star-forming galaxy and a starburst, in combination with a type-1 AGN template for which we allow for additional dust extinction (five free parameters). We fit the primary using the total FUV and NUV fluxes\footnote{Justified by the red optical SED of the secondary} and the deblended SDSS and UKIDSS photometry, but using the WISE fluxes as upper bounds. As expected, the residual FUV and NUV fluxes from the fit are consistent with 0; so for the secondary, we use the 3-$\sigma$ upper bounds in FUV and NUV based on their 1-$\sigma$ errors. The best-fit primary SED can only account for part of the WISE fluxes; so for the secondary, we use the residual fluxes in the WISE bands. The SED fits of both components are shown in Figure \ref{fg:sed_fit}.

The best-fit SEDs are consistent with the type-1$-$type-2 AGN pair scenario. The AGN in the primary is slightly reddened ($E(B-V)$ = 0.025). The secondary component requires a heavily reddened AGN ($E(B-V)$ = 4.2) to match the blue $3.4-4.6$\um\ color that is indicative of warm dust. In the following, we use the best-fit SED models to derive physical properties of the binary AGN. The quoted error-bars of the luminosity/mass estimates discussed in the following sections are obtained after scaling the flux error-bars to yield best-fit SEDs with $\chi^2_{\nu}=1$.

\subsubsection{Black Hole and Host Galaxy Mass}

The black hole mass of the primary type-1 AGN is log$(M_{\rm BH}/M_{\odot})$ = 8.06$\pm$0.24 with FWHM(\Hb) = $3800\pm100$ \kms\ and $\nu L_{\rm 5100}$ = $(9.5\pm3.0)\times10^{43}$ erg~s$^{-1}$ \citep[Eq. 5 of][]{Vestergaard06}. The \Hb\ FWHM is from the SDSS spectrum. The 5100\AA\ AGN luminosity is from the SED fit of the primary. The uncertainty in the black hole mass is dominated by the uncertainties in the virial coefficient and in the size-luminosity relation \citep{Kaspi05}. 

The best-fit SEDs imply stellar masses of $(1.7\pm0.3)\times10^{11}~\rm M_{\odot}$ and $(2.4\pm0.4)\times10^{11}~\rm M_{\odot}$ for the primary and secondary host galaxies, respectively, suggesting \obj\ is close to an equal-mass merger. The stellar masses are estimated using the color$-$mass-to-light-ratio relation of \citet{Bell03}. The black hole mass and the host galaxy mass of the primary AGN is consistent with the Magorrian relation \citep[$M_{\rm BH}/M_{\rm Bulge} \simeq$ 0.14\%;][]{Haring04,Magorrian98}. The Keck AO image indicates early-type morphologies for both host galaxies (\S~\ref{sec:morph}). 

\subsubsection{Black Hole Accretion Rate and Star Formation Rate}

We estimate black hole accretion rates from the AGN bolometric luminosities. \citet{Gallagher07} argue that monochromatic 3\um\ luminosity is a more robust indicator of bolometric AGN luminosity than the conventional 5100\AA\ luminosity because of its lack of sensitivity to dust extinction. The best-fit SEDs give AGN rest-frame 3\um\ luminosities of $\nu L_{\nu}$(3\um) = $(1.6\pm0.5)\times10^{44}$ and $(3.1\pm0.9)\times10^{44}$ erg~s$^{-1}$ for the primary and secondary, respectively. Adopting the bolometric corrections from \citet{Hopkins07}, we get $L_{\rm bol} = (1.7\pm0.5)\times10^{45}$ and $(3.3\pm1.0)\times10^{45}$ erg~s$^{-1}$. The total bolometric luminosity agrees well with the \rosat\ soft X-ray derived bolometric luminosity, $L_{\rm bol} = (5.2\pm1.8)\times10^{45}$ erg~s$^{-1}$, using also the SED templates of \citet{Hopkins07}. The corresponding accretion rates are $\dot{M}_{\rm BH}$ = $0.30\pm0.09$ and $0.58\pm0.17$ $(0.1/\epsilon)$ \msun~yr$^{-1}$, where $\epsilon$ is the radiative efficiency of the central engine. Assuming the same black hole to host galaxy mass ratio for the two components, the Eddington ratios are 0.12 and 0.16 for the primary and secondary, respectively, which are close to the mean Eddington ratios of optically selected AGNs at $z \sim 0.4$ \citep{McLure04}. 

Based on ultraviolet luminosities \citep{Kennicutt98}, the models give SFRs of $7\pm3$ and $8\pm1$ \msun~yr$^{-1}$ for the primary and secondary, respectively, consistent with \OII-derived upper limits (\S~\ref{sec:sf_vs_agn}).

\subsubsection{Radio Loudness}

Both AGNs in \obj\ are radio-moderate-to-loud. The rest-frame 5~GHz to 4400\AA\ flux ratio of the primary gives a radio-loudness of $R$ = 9.8$\pm$2.9, which is on the borderline between radio-moderate and radio-loud AGNs \citep{Kellermann89}. The secondary AGN's optical emission is heavily obscured, and the extinction-corrected 4400\AA\ flux of $0.18\pm0.05$ mJy implies $R$ = 12.3$\pm$3.6. 

\subsection{Host Galaxy Morphology} \label{sec:morph}

The primary appears essentially unresolved in the Keck AO image; it was used as the tip-tilt reference star for the LGSAO observations. There are no PSF reference stars within the field. However, we have found, in agreement with \citet{Carrasco10}, that the effective radius $r_e$ is not very sensitive to uncertainties in the PSF. We use a PSF derived from a well-exposed but unsaturated star in the field of another AGN, SDSS\,J131106.73+195234.2, which we observed on the same night. This star was the tip-tilt reference. It is slightly brighter (SDSS $r = 17.0$) than \obj\ and the AO-corrected image has FWHM = 0.10\arcsec. Since the primary of \obj\ clearly has an unresolved component, we use an iterative process to disentangle the morphologies. We first convolve the PSF star with a Gaussian to match its FWHM to that of the Gaussian fit to the primary of \obj, measured within a 2-pixel (0.08\arcsec) radius. We use this modified PSF to model both components: the primary as a combination of a PSF and a \sersic\ profile, and the secondary as a single \sersic\ profile. We then use the output of this model to update the PSF FWHM and iterate. The final PSF FWHM is about 0.15\arcsec.

The unresolved component comprises $12^{+8}_{-4}$\%\ of the total flux of the primary, roughly consistent with that from the SED fitting (24$\pm$9\%; \S~\ref{sec:sed}). The primary host galaxy is significantly more compact than the secondary, with $r_e = 3.7 \pm 0.2$ kpc, compared to the secondary's $r_e = 6.5 \pm 0.6$ kpc. The \sersic\ indices for the primary and secondary host galaxies are $n = 4.8 \pm 2.0$ and $2.9 \pm 0.2$, respectively. In spite of these differences, the masses and effective radii for both host galaxies are consistent with the mass$-$size relation of local early-type galaxies \citep{Damjanov09}.

\section{Summary}

\obj\ was initially identified as a candidate type-1$-$type-2 AGN pair by extensive AO imaging and integral-field spectroscopic followup of SDSS double-peaked AGNs. Here we have presented its EVLA images at 1.4, 5.0, and 8.5~GHz. These data reveal that each nucleus of the galaxies is associated with a steep spectrum compact radio source. The radio powers are inconsistent with star formation, indicating that the radio emission is powered by accretion. Therefore, the EVLA data confirm that \obj\ is a 7.4~kpc-separation type-1$-$type-2 AGN pair.

The AGNs appear to be normal Seyferts. SED fitting shows that \obj\ is a merger of two $\sim10^{11}$ \msun\ galaxies. The accretion rates are 0.30 and 0.58 $(0.1/\epsilon)$ \msun~yr$^{-1}$ for the primary and the secondary, respectively. With both black hole masses around $10^8$ \msun, the AGNs are accreting at Eddington ratios of $\sim$0.15. With an angular separation of 1.4\arcsec, future \chandra\ observations can provide spatially resolved X-ray imaging and spectroscopy of the binary. 

It is surprising that both AGNs are close to radio-loud ($R \sim 10$), because there are merely four binary AGN candidates from our survey (detection statistics consistent with that of \citealt{Tingay11}). Finding one such object in four requires an extremely high radio-loud fraction, $50_{-10}^{+27}$\%, for AGNs in kpc-scale binaries. A larger sample is clearly needed to further evaluate this result.

\acknowledgments
ZYZ acknowledges support by NSFC key project 10833006. ADM is a research fellow of the Alexander von Humboldt Foundation of Germany. AS and SGD were partially supported by NSF grants AST-0807900 and AST-0909182, respectively. SGD also acknowledges support by the Ajax Foundation. The National Radio Astronomy Observatory is a facility of the National Science Foundation operated under cooperative agreement by Associated Universities, Inc.

{\it Facilities}: VLA, Keck:II (LGSAO/NIRC2), UH2.2m (SNIFS), Sloan


\begin{thebibliography}
\expandafter\ifx\csname natexlab\endcsname\relax\def\natexlab#1{#1}\fi

\bibitem[{Aihara {et~al.}(2011)Aihara, Allende~Prieto, An, Anderson, Aubourg,
  Balbinot, Beers, Berlind, Bickerton, Bizyaev, Blanton, Bochanski, Bolton,
  Bovy, Brandt, Brinkmann, Brown, Brownstein, Busca, Campbell, Carr, Chen,
  Chiappini, Comparat, Connolly, Cortes, Croft, Cuesta, da~Costa, Davenport,
  Dawson, Dhital, Ealet, Ebelke, Edmondson, Eisenstein, Escoffier, Esposito,
  Evans, Fan, Femen{\'\i}a~Castell{\'a}, Font-Ribera, Frinchaboy, Ge,
  Gillespie, Gilmore, Gonz{\'a}lez~Hern{\'a}ndez, Gott, Gould, Grebel, Gunn,
  Hamilton, Harding, Harris, Hawley, Hearty, Ho, Hogg, Holtzman, Honscheid,
  Inada, Ivans, Jiang, Johnson, Jordan, Jordan, Kazin, Kirkby, Klaene, Knapp,
  Kneib, Kochanek, Koesterke, Kollmeier, Kron, Lampeitl, Lang, Le~Goff, Lee,
  Lin, Long, Loomis, Lucatello, Lundgren, Lupton, Ma, MacDonald, Mahadevan,
  Maia, Makler, Malanushenko, Malanushenko, Mandelbaum, Maraston, Margala,
  Masters, McBride, McGehee, McGreer, M{\'e}nard, Miralda-Escud{\'e}, Morrison,
  Mullally, Muna, Munn, Murayama, Myers, Naugle, Fausti~Neto, Cuong~Nguyen,
  Nichol, O'Connell, Ogando, Olmstead, Oravetz, Padmanabhan,
  Palanque-Delabrouille, Pan, Pandey, P{\^a}ris, Percival, Petitjean,
  Pfaffenberger, Pforr, Phleps, Pichon, Pieri, Prada, Price-Whelan, Raddick,
  Ramos, Reyl{\'e}, Rich, Richards, Rix, Robin, Rocha-Pinto, Rockosi, Roe,
  Rollinde, Ross, Ross, Rossetto, S{\'a}nchez, Sayres, Schlegel, Schlesinger,
  Schmidt, Schneider, Sheldon, Shu, Simmerer, Simmons, Sivarani, Snedden,
  Sobeck, Steinmetz, Strauss, Szalay, Tanaka, Thakar, Thomas, Tinker,
  Tofflemire, Tojeiro, Tremonti, Vandenberg, Vargas~Maga{\~n}a, Verde, Vogt,
  Wake, Wang, Weaver, Weinberg, White, White, Yanny, Yasuda, Yeche, \&
  Zehavi}]{Aihara11}
Aihara, H., {et~al.} 2011, \apjs, 193, 29

\bibitem[{Aldering {et~al.}(2002)Aldering, Adam, Antilogus, Astier, Bacon,
  Bongard, Bonnaud, Copin, Hardin, Henault, Howell, Lemonnier, Levy, Loken,
  Nugent, Pain, Pecontal, Pecontal, Perlmutter, Quimby, Schahmaneche, Smadja,
  \& Wood-Vasey}]{Aldering02}
Aldering, G., {et~al.} 2002, in SPIE Conference Series, Vol. 4836, 61--72

\bibitem[{Assef {et~al.}(2010)Assef, Kochanek, Brodwin, Cool, Forman, Gonzalez,
  Hickox, Jones, Le~Floc'h, Moustakas, Murray, \& Stern}]{Assef10}
Assef, R.~J., {et~al.} 2010, \apj, 713, 970

\bibitem[{Baldwin {et~al.}(1981)Baldwin, Phillips, \& Terlevich}]{Baldwin81}
Baldwin, J.~A., Phillips, M.~M., \& Terlevich, R. 1981, \pasp, 93, 5

\bibitem[{Barnes \& Hernquist(1996)}]{Barnes96}
Barnes, J.~E., \& Hernquist, L. 1996, \apj, 471, 115

\bibitem[{Bell {et~al.}(2003)Bell, McIntosh, Katz, \& Weinberg}]{Bell03}
Bell, E.~F., McIntosh, D.~H., Katz, N., \& Weinberg, M.~D. 2003, \apjs, 149,
  289

\bibitem[{Brusa {et~al.}(2007)Brusa, Zamorani, Comastri, Hasinger, Cappelluti,
  Civano, Finoguenov, Mainieri, Salvato, Vignali, Elvis, Fiore, Gilli, Impey,
  Lilly, Mignoli, Silverman, Trump, Urry, Bender, Capak, Huchra, Kneib,
  Koekemoer, Leauthaud, Lehmann, Massey, Matute, McCarthy, McCracken, Rhodes,
  Scoville, Taniguchi, \& Thompson}]{Brusa07}
Brusa, M., {et~al.} 2007, \apjs, 172, 353

\bibitem[Burke-Spolaor(2011)]{Burke11} Burke-Spolaor, S.\ 2011, \mnras, 410, 2113 

\bibitem[{Cardelli {et~al.}(1989)Cardelli, Clayton, \& Mathis}]{Cardelli89}
Cardelli, J.~A., Clayton, G.~C., \& Mathis, J.~S. 1989, \apj, 345, 245

\bibitem[{Carrasco {et~al.}(2010)Carrasco, Conselice, \& Trujillo}]{Carrasco10}
Carrasco, E.~R., Conselice, C.~J., \& Trujillo, I. 2010, \mnras, 405, 2253

\bibitem[{Comerford {et~al.}(2009)Comerford, Gerke, Newman, Davis, Yan, Cooper,
  Faber, Koo, Coil, Rosario, \& Dutton}]{Comerford09a}
Comerford, J.~M., {et~al.} 2009, \apj, 698, 956

\bibitem[{Damjanov {et~al.}(2009)Damjanov, McCarthy, Abraham, Glazebrook, Yan,
  Mentuch, Le~Borgne, Savaglio, Crampton, Murowinski, Juneau, Carlberg,
  J{\o}rgensen, Roth, Chen, \& Marzke}]{Damjanov09}
Damjanov, I., {et~al.} 2009, \apj, 695, 101

\bibitem[{Fu {et~al.}(2011{\natexlab{a}})Fu, Myers, Djorgovski, \& Yan}]{Fu11a}
Fu, H., Myers, A.~D., Djorgovski, S.~G., \& Yan, L. 2011{\natexlab{a}}, \apj,
  733, 103

\bibitem[{Fu {et~al.}(2011{\natexlab{b}})Fu, Yan, Myers, Stockton, Djorgovski,
  Aldering, \& Rich}]{Fu11b}
Fu, H., Yan, L., Myers, A.~D., Stockton, A., Djorgovski, S.~G., Aldering, G.,
  \& Rich, J.~A. 2011{\natexlab{b}}, preprint, arXiv:1107:3564

\bibitem[{Gallagher {et~al.}(2007)Gallagher, Richards, Lacy, Hines, Elitzur, \&
  Storrie-Lombardi}]{Gallagher07}
Gallagher, S.~C., Richards, G.~T., Lacy, M., Hines, D.~C., Elitzur, M., \&
  Storrie-Lombardi, L.~J. 2007, \apj, 661, 30

\bibitem[{Gerke {et~al.}(2007)Gerke, Newman, Lotz, Yan, Barmby, Coil,
  Conselice, Ivison, Lin, Koo, Nandra, Salim, Small, Weiner, Cooper, Davis,
  Faber, \& Guhathakurta}]{Gerke07}
Gerke, B.~F., {et~al.} 2007, \apj, 660, L23

\bibitem[{H{\"a}ring \& Rix(2004)}]{Haring04}
H{\"a}ring, N., \& Rix, H.-W. 2004, \apj, 604, L89

\bibitem[{Hennawi {et~al.}(2010)Hennawi, Myers, Shen, Strauss, Djorgovski, Fan,
  Glikman, Mahabal, Martin, Richards, Schneider, \& Shankar}]{Hennawi10}
Hennawi, J.~F., {et~al.} 2010, \apj, 719, 1672

\bibitem[{Ho(2008)}]{Ho08}
Ho, L.~C. 2008, \araa, 46, 475

\bibitem[{Hopkins {et~al.}(2007)Hopkins, Richards, \& Hernquist}]{Hopkins07}
Hopkins, P.~F., Richards, G.~T., \& Hernquist, L. 2007, \apj, 654, 731

\bibitem[{Junkkarinen {et~al.}(2001)Junkkarinen, Shields, Beaver, Burbidge,
  Cohen, Hamann, \& Lyons}]{Junkkarinen01}
Junkkarinen, V., Shields, G.~A., Beaver, E.~A., Burbidge, E.~M., Cohen, R.~D.,
  Hamann, F., \& Lyons, R.~W. 2001, The Astrophysical Journal Letters, 549,
  L155

\bibitem[{Kaspi {et~al.}(2005)Kaspi, Maoz, Netzer, Peterson, Vestergaard, \&
  Jannuzi}]{Kaspi05}
Kaspi, S., Maoz, D., Netzer, H., Peterson, B.~M., Vestergaard, M., \& Jannuzi,
  B.~T. 2005, \apj, 629, 61

\bibitem[{Kellermann {et~al.}(1989)Kellermann, Sramek, Schmidt, Shaffer, \&
  Green}]{Kellermann89}
Kellermann, K.~I., Sramek, R., Schmidt, M., Shaffer, D.~B., \& Green, R. 1989,
  \aj, 98, 1195

\bibitem[{Kennicutt(1998)}]{Kennicutt98}
Kennicutt, Robert~C., J. 1998, \araa, 36, 189

\bibitem[{Komossa {et~al.}(2003)Komossa, Burwitz, Hasinger, Predehl, Kaastra,
  \& Ikebe}]{Komossa03}
Komossa, S., Burwitz, V., Hasinger, G., Predehl, P., Kaastra, J.~S., \& Ikebe,
  Y. 2003, \apj, 582, L15

\bibitem[{Koss {et~al.}(2011)Koss, Mushotzky, Treister, Veilleux, Vasudevan,
  Miller, Sanders, Schawinski, \& Trippe}]{Koss11}
Koss, M., {et~al.} 2011, \apj, 735, L42

\bibitem[{Lawrence {et~al.}(2007)Lawrence, Warren, Almaini, Edge, Hambly,
  Jameson, Lucas, Casali, Adamson, Dye, Emerson, Foucaud, Hewett, Hirst,
  Hodgkin, Irwin, Lodieu, McMahon, Simpson, Smail, Mortlock, \&
  Folger}]{Lawrence07}
Lawrence, A., {et~al.} 2007, \mnras, 379, 1599

\bibitem[{Liu {et~al.}(2010{\natexlab{a}})Liu, Greene, Shen, \&
  Strauss}]{Liu10b}
Liu, X., Greene, J.~E., Shen, Y., \& Strauss, M.~A. 2010{\natexlab{a}}, \apj,
  715, L30

\bibitem[{Liu {et~al.}(2010{\natexlab{b}})Liu, Shen, Strauss, \&
  Greene}]{Liu10a}
Liu, X., Shen, Y., Strauss, M.~A., \& Greene, J.~E. 2010{\natexlab{b}}, \apj,
  708, 427

\bibitem[{Liu {et~al.}(2011)Liu, Shen, Strauss, \& Hao}]{Liu11a}
Liu, X., Shen, Y., Strauss, M.~A., \& Hao, L. 2011, preprint, arXiv:1104.0950

\bibitem[{Magorrian {et~al.}(1998)Magorrian, Tremaine, Richstone, Bender,
  Bower, Dressler, Faber, Gebhardt, Green, Grillmair, Kormendy, \&
  Lauer}]{Magorrian98}
Magorrian, J., {et~al.} 1998, \aj, 115, 2285

\bibitem[{Martin {et~al.}(2005)Martin, Fanson, Schiminovich, Morrissey,
  Friedman, Barlow, Conrow, Grange, Jelinsky, Milliard, Siegmund, Bianchi,
  Byun, Donas, Forster, Heckman, Lee, Madore, Malina, Neff, Rich, Small,
  Surber, Szalay, Welsh, \& Wyder}]{Martin05}
Martin, D.~C., {et~al.} 2005, \apjl, 619, L1

\bibitem[{McGurk {et~al.}(2011)McGurk, Max, Rosario, Shields, Smith, \&
  Wright}]{McGurk11}
McGurk, R.~C., Max, C.~E., Rosario, D.~J., Shields, G.~A., Smith, K.~L., \&
  Wright, S.~A. 2011, arXiv:1107.2651

\bibitem[{McLure \& Dunlop(2004)}]{McLure04}
McLure, R.~J., \& Dunlop, J.~S. 2004, \mnras, 352, 1390

\bibitem[{Myers {et~al.}(2008)Myers, Richards, Brunner, Schneider, Strand,
  Hall, Blomquist, \& York}]{Myers08}
Myers, A.~D., Richards, G.~T., Brunner, R.~J., Schneider, D.~P., Strand, N.~E.,
  Hall, P.~B., Blomquist, J.~A., \& York, D.~G. 2008, \apj, 678, 635

\bibitem[O'Dea(1998)]{ODea98} O'Dea, C.~P.\ 1998, \pasp, 110, 493 

\bibitem[{Osterbrock \& Ferland(2006)}]{Osterbrock06}
Osterbrock, D.~E., \& Ferland, G.~J. 2006, Astrophysics of gaseous nebulae and
  active galactic nuclei (Sausalito, CA: University Science Books, 2006)

\bibitem[{Peng {et~al.}(2010)Peng, Ho, Impey, \& Rix}]{Peng10}
Peng, C.~Y., Ho, L.~C., Impey, C.~D., \& Rix, H.-W. 2010, \aj, 139, 2097

\bibitem[{Perley {et~al.}(2011)Perley, Chandler, Butler, \& Wrobel}]{Perley11}
Perley, R.~A., Chandler, C.~J., Butler, B.~J., \& Wrobel, J.~M. 2011,
  arXiv:1106.0532

\bibitem[{Richstone {et~al.}(1998)Richstone, Ajhar, Bender, Bower, Dressler,
  Faber, Filippenko, Gebhardt, Green, Ho, Kormendy, Lauer, Magorrian, \&
  Tremaine}]{Richstone98}
Richstone, D., {et~al.} 1998, Nature, 395, 14

\bibitem[{Shen {et~al.}(2011)Shen, Liu, Greene, \& Strauss}]{Shen11}
Shen, Y., Liu, X., Greene, J.~E., \& Strauss, M.~A. 2011, \apj, 735, 48

\bibitem[{Smith {et~al.}(2010)Smith, Shields, Bonning, McMullen, Rosario, \&
  Salviander}]{Smith10}
Smith, K.~L., Shields, G.~A., Bonning, E.~W., McMullen, C.~C., Rosario, D.~J.,
  \& Salviander, S. 2010, \apj, 716, 866
  
\bibitem[Tingay \& Wayth(2011)]{Tingay11} Tingay, S.~J., \& Wayth, R.~B.\ 2011, \aj, 141, 174 

\bibitem[{Vestergaard \& Peterson(2006)}]{Vestergaard06}
Vestergaard, M., \& Peterson, B.~M. 2006, \apj, 641, 689

\bibitem[{Wang {et~al.}(2009)Wang, Chen, Hu, Mao, Zhang, \& Bian}]{Wang09}
Wang, J.-M., Chen, Y.-M., Hu, C., Mao, W.-M., Zhang, S., \& Bian, W.-H. 2009,
  \apj, 705, L76

\bibitem[{White {et~al.}(1997)White, Becker, Helfand, \& Gregg}]{White97}
White, R.~L., Becker, R.~H., Helfand, D.~J., \& Gregg, M.~D. 1997, \apj, 475,
  479

\bibitem[{Wizinowich {et~al.}(2006)Wizinowich, Le~Mignant, Bouchez, Campbell,
  Chin, Contos, van Dam, Hartman, Johansson, Lafon, Lewis, Stomski, Summers,
  Brown, Danforth, Max, \& Pennington}]{Wizinowich06}
Wizinowich, P.~L., {et~al.} 2006, \pasp, 118, 297

\bibitem[{Wright {et~al.}(2010)Wright, Eisenhardt, Mainzer, Ressler, Cutri,
  Jarrett, Kirkpatrick, Padgett, McMillan, Skrutskie, Stanford, Cohen, Walker,
  Mather, Leisawitz, Gautier, McLean, Benford, Lonsdale, Blain, Mendez, Irace,
  Duval, Liu, Royer, Heinrichsen, Howard, Shannon, Kendall, Walsh, Larsen,
  Cardon, Schick, Schwalm, Abid, Fabinsky, Naes, \& Tsai}]{Wright10}
Wright, E.~L., {et~al.} 2010, \aj, 140, 1868

\bibitem[{Yun {et~al.}(2001)Yun, Reddy, \& Condon}]{Yun01}
Yun, M.~S., Reddy, N.~A., \& Condon, J.~J. 2001, \apj, 554, 803

\end{thebibliography}
\end{document}